\documentstyle[12pt,aps,epsfig]{revtex}

\begin{document}
\draft
\title{Single Spin Asymmetry in Heavy Quark\\ 
Photoproduction (and Decay) as a Test of pQCD\footnote{Talk given at the 
NATO Advanced Spin Physics Workshop,  Yerevan, June 30- July 3, 2002. }}
\author{N.Ya.Ivanov\footnote{E-mail: nikiv@uniphi.yerphi.am}}
\address{Yerevan Physics Institute,  Alikhanian Br.2, 375036 Yerevan, Armenia}
\maketitle\vspace{3mm}

\begin{abstract}
We review the properties of the single spin asymmetry (SSA) in heavy quark production 
by linearly polarized photons and analyze the possibility to measure this SSA in the 
planned E160/E161 experiments at SLAC.
\end{abstract}

\section{Introduction}
In the framework of perturbative QCD, the basic spin-averaged
characteristics of heavy flavor hadro-, photo- and
electroproduction are known exactly up to the next-to-leading
order (NLO). During the last ten years, these NLO results have
been widely used for a phenomenological description of available
data (for a review see \cite{1}). At the same time, the key
question still remains open: How to test the applicability of QCD
at fixed order to the heavy quark production? The problem is
twofold. On the one hand, the NLO corrections are large; they
increase the leading order (LO) predictions for both charm and
bottom production cross sections approximately by a factor of two.
For this reason, one could expect that higher-order corrections,
as well as nonperturbative contributions, can be essential in
these processes, especially for the $c$-quark case. On the other
hand, it is very difficult to compare directly, without additional
assumptions, pQCD predictions for spin-averaged cross sections
with experimental data because of a high sensivity of the
theoretical calculations to standard uncertainties in the input
QCD parameters: $m_{Q}$, the factorization and renormalization scales, 
$\mu _{F}$ and  $\mu _{R}$, $\Lambda_{QCD}$ and the parton distribution 
functions \cite{2,3}.

In recent years, the role of higher-order corrections has been
extensively investigated in the framework of the soft gluon
resummation formalism \cite{5,6,7}. Unfortunately, formally
resummed cross sections are ill-defined due to the Landau pole
contribution, and numerical predictions for the heavy quark 
production cross sections can depend significantly on the choice 
of resummation prescription \cite{8}. Another open question, also closely related 
to convergence of the perturbative series, is the role of subleading 
contributions which are not, in principle, under control of the
resummation procedure \cite{8,8L}.

For this reason, it is of special interest to study those
observables which are well defined in pQCD. A nontrivial example
of such an observable is proposed in  \cite{9,10}, where the charm
and bottom production by linearly polarized photons,
\begin{equation}
\gamma ^{\uparrow }+N\rightarrow Q+X[\overline{Q}],  \label{1}
\end{equation}
was considered\footnote{The well known examples are the shapes of
differential cross sections of heavy flavor production which are
sufficiently stable under radiative corrections.}. In particular,
the single spin asymmetry (SSA) parameter, $A_{Q}(p_{QT})$, which
measures the parallel-perpendicular asymmetry in the quark
azimuthal distribution,
\begin{equation}
\frac{{{\rm{d}}}^{2}\sigma _{Q}}{{{\rm{d}}}p_{QT}{{\rm{d}}}\varphi _{Q}}
(p_{QT},\varphi _{Q})=\frac{1}{2\pi }\frac{{{\rm{d}}}\sigma _{Q}^{\rm{unp}}}
{{{\rm{d}}}p_{QT}}(p_{QT})\left[ 1+A_{Q}(p_{QT}){\cal P}_{\gamma }\cos
2\varphi _{Q}\right],   \label{2}
\end{equation}
has been calculated. In (\ref{2}) $\frac{{\rm{d}}\sigma _{Q}^{
\rm{unp}}}{{\rm{d}}p_{QT}}$ is the unpolarized cross section,${\cal P}%
_{\gamma }$ is the degree of linear polarization of the incident
photon beam and $\varphi _{Q}$ is the angle between the beam
polarization direction and the observed quark transverse momentum,
$p_{QT}$. The following remarkable properties of the SSA,
$A_{Q}(p_{QT})$, have been observed  \cite{9}:

\begin{itemize}
\item  The azimuthal asymmetry (\ref{2}) is of leading twist; in a
wide kinematical region, it is predicted to be about $0.2$ for
both charm and bottom quark production.
\item  At energies sufficiently above the production threshold, the
LO predictions for $A_{Q}(p_{QT})$ are insensitive (to within few
percent) to uncertainties in the QCD input parameters.
\item   Nonperturbative corrections to the $b$-quark azimuthal asymmetry
are negligible. Because of the smallness of the $c$-quark mass,
the analogous corrections to $A_{c}(p_{QT})$ are larger; they are
of the order of $20\%$.
\end{itemize}

In Ref. \cite{10}, radiative corrections to the $\varphi-$dependent 
cross section (\ref{2}) have been investigated in the soft-gluon 
approximation. Calculations \cite{10} indicate a high perturbative
stability of the pQCD predictions for $A_{Q}$. In particular,

\begin{itemize}
\item  At the next-to-leading logarithmic (NLL) level, the NLO and NNLO
predictions for $A_{Q}$ affect the LO results by less than
$1\%$ and $2\%$, respectively.
\item  Computations of the higher order contributions (up to the 6th order
in $\alpha _{s}$) to the NLL accuracy lead only to a few percent
corrections to the Born result for $A_{Q}$. This implies that
large soft-gluon contributions to the spin-dependent and
unpolarized cross sections cancel each other  with a
good accuracy.
\end{itemize}

So, contrary to the the production cross sections, the single spin
asymmetry in heavy flavor photoproduction is an observable
quantitatively well defined in pQCD: it is stable, both
parametricaly and perturbatively, and insensitive to
nonperturbative corrections. Measurements of this asymmetry would
provide an ideal test of the conventional parton model based on
pQCD.

Concerning the experimental aspects, the azimuthal asymmetry in
charm photoproduction can be measured at SLAC where a coherent
bremsstrahlung beam of linearly polarized photons with energies up
to 40 GeV will be available soon \cite{11}. In the planned E160
and E161 experiments, the charm production will be investigated
using the inclusive spectra of the decay lepton:
\begin{equation}
\gamma ^{\uparrow }+N\rightarrow c+X[\overline{c}]\rightarrow \mu ^{+}+X.
\label{4}
\end{equation}

In this paper, we analyze a possibility to measure the SSA in heavy quark
photoproduction using the decay lepton spectra. We calculate the SSA in the
decay lepton azimuthal distribution:
\begin{equation}
\frac{{\rm{d}}^{2}\sigma _{\ell }}{{\rm{d}}p_{\ell T}{\rm{d}}\varphi _{\ell }%
}(p_{\ell T},\varphi _{\ell })=\frac{1}{2\pi }\frac{{\rm{d}}\sigma _{\ell }^{%
{\rm{unp}}}}{{\rm{d}}p_{\ell T}}(p_{\ell T})\left[ 1+A_{\ell
}(p_{\ell T}){\cal P}_{\gamma }\cos 2\varphi _{\ell }\right] ,  \label{5}
\end{equation}
where $\varphi _{\ell }$ is the angle between the photon polarization
direction and the decay lepton transverse momentum, $p_{\ell T}$. Our main
results can be formulated as follows \cite{we}:

\begin{itemize}
\item  The SSA transferred from the decaying $c$-quark to the decay muon is
large in the SLAC kinematics; the ratio
$A_{\ell}(p_{T})/A_{c}(p_{T})$ is about $90\%$ for $p_{T}>$1 GeV.
\item  pQCD predictions for $A_{\ell }(p_{\ell T})$ are also stable, both
perturbatively and parametricaly.
\item  Nonperturbative corrections to $A_{\ell }(p_{\ell T})$ due to the gluon
transverse motion in the target and the $c$-quark fragmentation
are small; they are about $10\%$ for $p_{\ell T}>$1 GeV.
\item  The SSA (\ref{5}) depends weekly on theoretical uncertainties in
the charm semileptonic decays\footnote{For a review see Ref.
\cite{21}.}, $c\rightarrow \ell ^{+}\nu _{\ell }X_{q}$ $(q=d,s)$.
In particular,
\begin{itemize}
\item Contrary to the the production cross sections, the asymmetry
$A_{\ell }(p_{\ell T})$ is practically insensitive to the
unobserved stange quark mass, $m_s$, for $p_{\ell T}>$1 GeV.
\item The bound state effects due to the Fermi motion of the $c$-quark inside
the $D$-meson have only a small impact on $A_{\ell }(p_{\ell T})$,
in practically the whole region of $p_{\ell T}$.
\end{itemize}
\end{itemize}

So, we conclude that the SSA in the decay lepton azimuthal distribution (\ref
{5}) is also well defined in the framework of perturbation theory and can be
used as a good test of pQCD applicability to heavy flavor production.

\section{pQCD Predictions for SSA}
\subsection{LO Results}
At the Born level, the only partonic subprocess which is
responsible for the reaction (\ref{4}) is the heavy quark
production in the photon-gluon fusion,
\begin{equation}
\gamma ^{\uparrow }(k_{\gamma })+g(k_{g})\rightarrow Q(p_{Q})+\overline{Q}%
(p_{\overline{Q}})\rightarrow \ell (p_{\ell })+\nu _{\ell }+q+\overline{Q},
\label{6}
\end{equation}
with subsequent decay $c\rightarrow \ell ^{+}\nu _{\ell }q$
$(q=d,s)$ in the charm case and $b\rightarrow \ell
^{-}\overline{\nu }_{\ell }q$ $(q=u,c)$ in the bottom one. To
calculate distributions of final particles appearing in a process
of production and subsequent decay, it is useful to adopt the
narrow-width approximation,
\begin{equation}
\frac{1}{\left( p_{Q}^{2}-m_{Q}^{2}\right) ^{2}+\Gamma _{Q}^{2}m_{Q}^{2}}%
\rightarrow \frac{\pi }{\Gamma _{Q}m_{Q}}\delta \left(
p_{Q}^{2}-m_{Q}^{2}\right),  \label{7}
\end{equation}
with $\Gamma _{Q}$ the total width of the heavy quark. Corrections
to this approximation are negligibly small in both charm and
bottom cases since they have a relative size ${\cal O}(\Gamma
_{Q}/m_{Q})$.

In the case of the linearly polarized photon, the heavy quark
produced in the reaction (\ref{6}) is unpolarized. For this
reason, the single-inclusive cross section for the decay lepton
production in (\ref{6}) is a simple convolution:
\begin{equation}
E_{\ell }\frac{{\rm{d}}^{3}\hat{\sigma}_{\ell }}{{\rm{d}}^{3}p_{\ell }}({%
\vec{p}}_{\ell })=\frac{1}{\Gamma _{Q}}\int \frac{{\rm{d}}^{3}p_{Q}}{E_{Q}}%
\frac{E_{Q}{\rm{d}}^{3}\hat{\sigma}_{Q}}{{\rm{d}}^{3}p_{Q}}({\vec{p}}_{Q})%
\frac{E_{\ell }{\rm{d}}^{3}\Gamma _{\rm{sl}}}{{\rm{d}}^{3}p_{\ell }}%
(p_{\ell }\cdot p_{Q}).  \label{8}
\end{equation}

The leading order predictions for the
$\varphi_{Q}$-dependent cross section of heavy flavor production,
\begin{eqnarray}
\frac{E_{Q}{\rm{d}}^{3}\hat{\sigma}_{Q}}{{\rm{d}}^{3}p_{Q}}({\vec{p}}_{Q})
&\equiv &\frac{2s{\rm{d}}^{3}\hat{\sigma}_{Q}}{{\rm{d}}u_{1}{\rm{d}}t_{1}%
{\rm{d}}\varphi _{Q}}\left( s,t_{1},u_{1},\varphi _{Q}\right)  \nonumber \\
&=&\frac{1}{\pi s}\left[ B_{Q}\left( s,t_{1},u_{1}\right) +\Delta
B_{Q}\left( s,t_{1},u_{1}\right) {\cal P}_{\gamma }\cos 2\varphi
_{Q}\right],  \label{9}
\end{eqnarray}
are given in \cite{9}. Radiative corrections to the cross section (\ref{9})
was investigated in the soft gluon approximation in Ref. \cite{10}.

At the tree level, the invariant width of the semileptonic decay 
$c\rightarrow \ell ^{+}\nu _{\ell }q$ $(q=d,s)$ can be written as
\begin{equation}
\frac{E_{\ell }{\rm{d}}^{3}\Gamma _{\rm{sl}}}{{\rm{d}}^{3}p_{\ell }}%
(x)\equiv I_{\rm{sl}}(x)=\frac{G_{F}^{2}m_{Q}^{3}}{(2\pi )^{4}}
\left| V_{CKM}\right| ^{2}\frac{x%
\left( 1-x-\delta ^{2}\right) ^{2}}{1-x} \label{13}
\end{equation}
Here $V_{CKM}$ denotes the corresponding element of the
Cabbibo-Kobayashi--Maskawa matrix, $G_{F}$ is the Fermi constant, 
$x=2(p_{\ell }\cdot p_{Q})/m_{Q}^{2}$ and $\delta=m_{q}/m_{Q}$.

Let us discuss the hadron level pQCD predictions for the asymmetry in
azimuthal distribution of the decay lepton. In this paper, we will consider
only the charm photoproduction at the SLAC\ energy $E_{\gamma }\approx$ 35 GeV, $%
E_{\gamma }=(S-m_{N}^{2})/2m_{N}$. Unless otherwise stated, the CTEQ5M \cite{12}
parametrization of the gluon distribution function is used. The default
value of the charm quark mass is $m_{c}=$ 1.5 GeV.

\begin{figure}
\begin{center}
\mbox{\epsfig{file=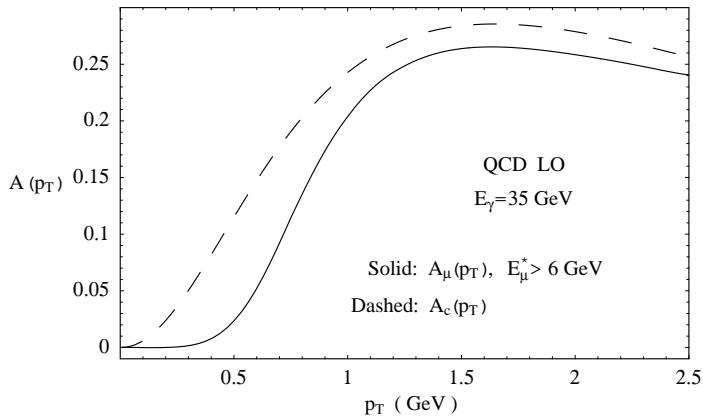,width=270pt}}
\caption{\label{Fig.1}
Comparison of the QCD LO pedictions for $A_{\mu }(p_{T})$ and
 $A_{c}(p_{T})$.}
\end{center}
\end{figure}

Our calculations of the quantities $A_{\mu }(p_{T})$ and
$A_{c}(p_{T})$ are given in Fig.\ref{Fig.1} by solid and dashed
lines, respectively. One can see that the asymmetry transferred
from the decaying $c$-quark to the decay muon is large in the SLAC
kinematics; the ratio $A_{\mu }(p_{T})/A_{c}(p_{T})$ is about
$90\%$ for $p_{T}>$1 GeV. Note that $p_{T}\equiv p_{QT}$ when we
consider the heavy quark production and $p_{T}\equiv p_{\ell T}$
when the quantity $A_{\mu }(p_{\ell T})$ is discussed.

We have analyzed also the dependence of the SSA in the lepton
distribution on the unobserved strange quark mass, $m_{s}$. Our analysis 
shows that the LO predictions for $A_{\mu}(p_{T})$ are practically 
independent of $\delta=m_s/m_c$ at sufficiently large $p_{T}>$1 GeV.
For more details see Ref. \cite{we}.

\subsection{Radiative Corrections}

We have computed both
spin-dependent and unpolarized differential distributions
(\ref{9}) of the heavy-quark photoproduction at NLO to the
next-to-leading logarithmic accuracy. The NLO corrections to the
width of the heavy quark semileptonic decays are known exactly
\cite{A1,A2}. We have found that radiative corrections to the
leptonic SSA, $A_{\mu }(p_{T})$, in the reaction (\ref{4}) are of
the order of (1-2)$\%$ in the SLAC kinematics.

Two main reasons are responsible for perturbative stability of the
quantity $A_{\mu }(p_{T})$. First, radiative corrections to the
SSA in heavy quark production are small \cite{10}. Second, the
ratio $I_{\rm{sl}}^{\rm{NLO}}(x)/ I_{\rm{sl}}^{\rm{Born}}(x)$ is a
constant practically at all $x$, except for a narrow endpoint
region $x\approx 1$ \cite{A2}. (Note that
$I_{\rm{sl}}^{\rm{Born}}(x)$ is the LO invariant width of the
semileptonic decay $c\rightarrow \ell ^{+}\nu _{\ell }X_{q}$ given
by (\ref{13}) while $I_{\rm{sl}}^{\rm{NLO}}(x)$ is the
corresponding NLO one.)

\section{Nonperturbative Contributions}

Let us discuss how the pQCD predictions for single spin asymmetry
are affected by nonperturbative contributions due to the intrinsic
transverse motion of the gluon and the hadronization of the
produced heavy quark. Because of the low $c$-quark mass, these
contributions are especially important in the description of the
cross section for charmed particle production \cite{1}. At the
same time, our analysis shows that nonperturbative corrections to
the single spin asymmetry are not large.

Hadronization effects in heavy flavor production are usually modeled with
the help of the Peterson fragmentation function \cite{15},
\begin{equation}
D(y)=\frac{a_{\varepsilon }}{y\left[ 1-1/y-\varepsilon /(1-y)\right] ^{2}},
\label{28}
\end{equation}
where $a_{\varepsilon }$ is a normalization factor and
$\varepsilon _{D}=0.06$ in the case of a $D$-meson production. 

\begin{figure}
\begin{center}
\mbox{\epsfig{file=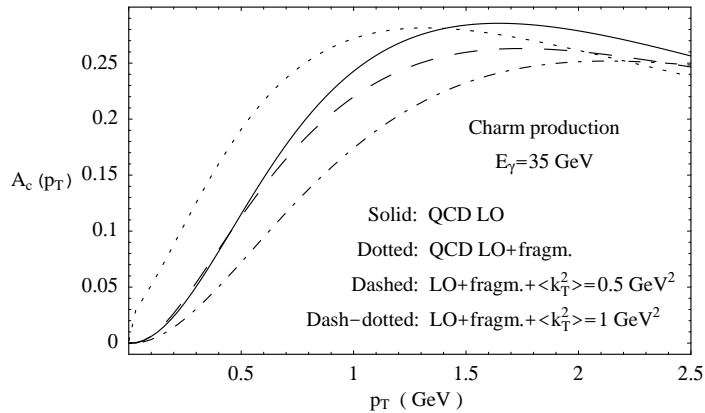,width=270pt}}
\end{center}
\caption{\label{Fig.3}
SSA in a $D$-meson production; the QCD LO predictions with
and without the inclusion of the $k_{T}$ smearing and Peterson
fragmentation effects.}
\end{figure}

Our calculations of the asymmetry in a $D$-meson production at LO
with and without the Peterson fragmentation effect are presented
in Fig.\ref{Fig.3} by dotted and solid curves, respectively. It is
seen that at $ p_{DT}\ge 1$ GeV the fragmentation corrections to
$A_{c}(p_{T})$ are less than $10\%$.

\begin{figure}
\begin{center}
\mbox{\epsfig{file=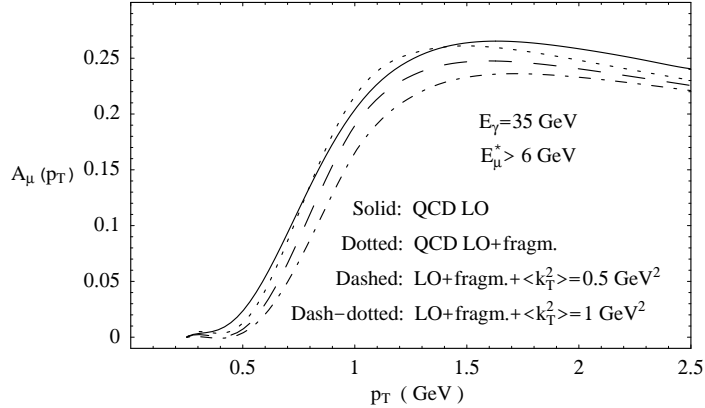,width=270pt}}
\end{center}
\caption{\label{Fig.4} SSA, $A_{\mu }(p_{T})$, in the decay lepton
distribution; the QCD LO predictions with and without the
inclusion of the $k_{T}$ smearing and Peterson fragmentation
effects.}
\end{figure}

Analogous corrections to the asymmetry in the decay lepton
azimuthal distribution, $A_{\mu }(p_{T}),$ are given in
Fig.\ref{Fig.4}. One can see that the effect of the fragmentation
function (\ref{28}) is practically negligible in the whole region
of $p_{\ell T}$.

To introduce $k_{T}$ degrees of freedom, $\vec{k}_{g}\simeq z\vec{k}_{N}+%
\vec{k}_{T}$, one extends the integral over the parton distribution function 
to the $k_{T}$-space,
\begin{equation}
{\rm{d}}zg(z,\mu _{F})\rightarrow {\rm{d}}z{\rm{d}}^{2}k_{T}f\left( \vec{k}%
_{T}\right) g(z,\mu _{F}).  \label{30}
\end{equation}
The transverse momentum distribution, $f\left( \vec{k}_{T}\right) $, is
usually taken to be a Gaussian:
\begin{equation}
f\left( \vec{k}_{T}\right) =\frac{{\rm{e}}^{-k_{T}^{2}/\langle
k_{T}^{2}\rangle }}{\pi \langle k_{T}^{2}\rangle }.  \label{31}
\end{equation}

Values of the $k_{T}$-kick corrections to the asymmetry in the
charm production, $A_{c}(p_{T})$, are shown in Fig.\ref{Fig.3} by
dashed ($\langle k_{T}^{2}\rangle =0.5$ GeV$^{2}$) and dash-dotted
($\langle k_{T}^{2}\rangle =1$ GeV$^{2}$) curves. One can see that
$k_{T}$-smearing is important only in the region of relatively low
$p_{QT}\leq m_{c}$. Note also that the fragmentation and
$k_{T}$-kick effects practically cancel each other in the case of
$\langle k_{T}^{2}\rangle =0.5$ GeV$^{2}$.

Corresponding calculations for the case of the lepton asymmetry
are presented in Fig.\ref{Fig.4}. It is seen that $A_{\mu
}(p_{T})$ is affected by $k_{T}$-corrections systematically
smaller than $A_{c}(p_{T})$.

\section{Conclusion}
In this paper we analyze the possibility to measure the SSA in
open charm photoproduction in the E160/E161 experiments at SLAC
where a coherent bremsstrahlung beam of linearly polarized photons
with energies up to 40 GeV will be available soon. In these
experiments, the charm production will be investigated with the
help of inclusive spectra of the secondary leptons. The SSA
transferred from the decaying $c$-quark to the decay muon is
predicted to be large for SLAC kinematics; the ratio $A_{\ell
}(p_{T})/A_{c}(p_{T})$ is about $90\%$ at $p_{T}>$1 GeV. Our
calculations show that the SSA in decay lepton distribution
preserves all remarkable properties of the SSA in heavy flavor
production: it is stable, both perturbatively and parametricaly,
and practically insensitive to nonperturbative contributions due
to the gluon transverse motion in the target and the heavy quark
fragmentation. We have also found that the QCD predictions for
$A_{\ell}(p_{T})$ depend weekly on theoretical uncertainties in
the charm semileptonic decays. We conclude that measurements of
$A_{\ell}(p_{T})$ in the E160/E161 experiments would provide a
good test of pQCD applicability to open charm production.

{\em Acknowledgements.} We would like to thank S.J. Brodsky, L. Dixon, 
A. Kotzinian, A.E. Kuraev, A.G. Oganesian, M.E. Peskin, A.V. Radyushkin and 
J. Tjon  for useful discussions.

\end{document}